\documentclass[a4paper]{jpconf}
\usepackage{oldgerm}
\usepackage{graphicx}
\def\half{{\textstyle\frac12}}

\def\fC{\textfrak{C}}
\def\fV{\textfrak{V}}
\def\fx{\textfrak{x}}
\def\fX{\textfrak{X}}
\begin{document}
\title{Gravity from Breaking of Local Lorentz Symmetry}

\author{Robertus Potting}

\address{CENTRA and Physics Department\\
Faculdade de Ci\^encias e Tecnologia\\
Universidade do Algarve\\
Faro, Portugal}

\ead{rpotting@ualg.pt}

\begin{abstract}
We present a model of gravity based on spontaneous Lorentz symmetry
breaking. We start from a model with spontaneously broken symmetries
for a massless 2-tensor with a linear kinetic term and a nonderivative
potential, which is shown to be equivalent to linearized general
relativity, with the Nambu-Goldstone (NG) bosons playing the role of
the gravitons. We apply a bootstrap procedure to the model based
on the principle of consistent coupling to the total energy
energy-momentum tensor. Demanding consistent application of the
bootstrap to the potential term severely restricts the form of the latter.
Nevertheless, suitable potentials exists that permit stable vacua.
It is shown that the resulting model
is equivalent, at low energy, to General Relativity in a fixed gauge.
\end{abstract}

\section{Symmetry vs.\ Broken Symmetry}
Masslessness  often arises as a consequence of the existence of a symmetry. In
quantum electrodynamics the masslessness of the photon is normally
attributed to gauge invariance, or symmetry under local changes of
phase. In quantum chromodynamics, the theory of the strong
interaction, masslessness of the gluons is likewise attributed to a
gauge invariance, albeit a nonlinear one. In general relativity, the
masslessness of gravitons can be traced to symmetry under active
diffeomorphisms: no diffeomorphism-invariant mass term exists.

In some circumstances, however, there exists an alternative reason why a
field might be massless. Surprisingly, this alternative explanation
involves the breaking of a symmetry rather than its existence. A
general result, the Nambu-Goldstone theorem~\cite{ng}, states under mild
assumptions that there must be a massless particle whenever a
continuous global symmetry of an action isn't a symmetry of the
vacuum.

In this talk, based on ongoing work with Alan Kosteleck\'y~\cite{kopo1,kopo2},
we show that an alternative description of gravity can
be constructed from a symmetric two-tensor without the assumption of
masslessness. In this picture, masslessness is a consequence of
symmetry breaking rather than of exact symmetry: diffeomorphism
symmetry and local Lorentz symmetry are spontaneously broken, but the
graviton remains massless because it is a Nambu-Goldstone mode.

The cardinal object in the theory is a symmetric two-tensor-density,
denoted by $C^{\mu\nu}$. Starting point is the Lagrange density
\begin{equation}
L=\half C^{\mu\nu}K_{\mu\nu\alpha\beta}C^{\alpha\beta} + 
V(C^{\mu\nu},\eta_{\mu\nu}).
\end{equation}
Here, $K_{\mu\nu\alpha\beta}$ is the usual quadratic kinetic operator
for a massless spin-2 field,
\begin{eqnarray}
K_{\mu\nu\alpha\beta}&=&\half[(-\eta_{\mu\nu}\eta_{\alpha\beta}
+\half\eta_{\mu\alpha}\eta_{\nu\beta}+\half\eta_{\mu\beta}\eta_{\nu\alpha})
\partial^2\nonumber\\
&&\quad{}+\eta_{\mu\nu}\partial_\alpha\partial_\beta
+\eta_{\alpha\beta}\partial_\mu\partial_\nu
-\half\eta_{\mu\alpha}\partial_\nu \partial_\beta
-\half\eta_{\nu\alpha}\partial_\mu \partial_\beta
-\half\eta_{\mu\beta}\partial_\nu\partial_\alpha
-\half\eta_{\nu\beta}\partial_\mu \partial_\alpha
],
\end{eqnarray}
and $V$ is a potential which is built out of the four scalars
\begin{eqnarray}
X_1&=&C^{\mu\nu}\eta_{\nu\mu},\label{X1}\\
X_2&=&(C\cdot\eta\cdot C\cdot\eta)^\mu_\mu,\label{X2}\\
X_3&=&(C\cdot\eta\cdot C\cdot\eta\cdot C\cdot\eta)^\mu_\mu,\label{X3}\\
X_4&=&(C\cdot\eta\cdot C\cdot\eta\cdot C\cdot\eta\cdot C\cdot\eta)^\mu_\mu.
\label{X4}
\end{eqnarray}
We will suppose that $V$ has a local minimum at
$C^{\mu\nu}=c^{\mu\nu}\ne0$, and that $C^{\mu\nu}$ acquires an
expectation value $\langle C^{\mu\nu}\rangle=c^{\mu\nu}$.
Note that this implies spontaneous breaking of Lorentz invariance.
We can now decompose $C^{\mu\nu}=c^{\mu\nu}+\tilde C^{\mu\nu}$ where
$\tilde C^{\mu\nu}$ are the fluctuations of $C^{\mu\nu}$.

At low energy, the values of the scalars $X_n$ will be approximately
fixed to their values $x_n$ in the local minimum. Then, the linearized
form of the potential can be taken equivalent to a sum of
Lagrange multiplier terms that fix the values of the four 
scalars $X_n$ to the values $x_n$:
\begin{equation}
V\to\sum_{n=1}^4\frac{\lambda_n}{n}(X_n-x_n).
\end{equation}
We obtain the linearized equations of motion
\begin{equation}
K_{\mu\nu\alpha\beta} \tilde C^{\alpha\beta}=
-\lambda_1\eta_{\mu\nu}+\lambda_2 (\eta c\eta)_{\mu\nu}+
\lambda_3 (\eta c\eta c\eta)_{\mu\nu}
+\lambda_4(\eta c\eta c\eta c\eta)_{\mu\nu}
\label{lin_eom}
\end{equation}
together with the constraints
\begin{equation}
\tilde C^\mu_\mu=0,\qquad
c^{\mu\nu}\tilde C_{\mu\nu}=0,\qquad
(c\eta c)^{\mu\nu}\tilde C_{\mu\nu}=0,\qquad
(c\eta c\eta c)^{\mu\nu}\tilde C_{\mu\nu}=0.
\label{gauge_conditions}
\end{equation}
Noting that the left-hand side of eq.\ (\ref{lin_eom}) equals the linear
part of the Ricci tensor, it follows that the low-energy linearized dynamics
of this model is equivalent to linearized general relativity in an axial-type
gauge defined by conditions (\ref{gauge_conditions}).

The four gauge conditions (\ref{gauge_conditions}) reduce the original
10 $h_{\mu\nu}$ modes to 6 degrees of freedom. They can be expressed as
the generators of the Lorentz generators $\mathcal{E}_{\mu\nu}=
-\mathcal{E}_{\mu\nu}$
acting on the vacuum expectation value $c^{\mu\nu}$ of the cardinal field:
\begin{equation}
\tilde C^{\mu\nu}=\mathcal{E}^\mu{}_\alpha c^{\alpha\nu}+
\mathcal{E}^\nu{}_\alpha c^{\mu\alpha}.
\end{equation}
Imposing the equations of motion imposes masslessness as well as the Lorenz
conditions:
\begin{equation}
\partial^2 \tilde C_{\mu\nu}=0,\qquad \partial^\mu \tilde C_{\mu\nu}=0,
\end{equation}
reducing the number of propagating degrees of freedom to two helicities.

\section{The bootstrap}

In order that the cardinal model correctly describe gravity it needs to
be coupled to the matter energy-momentum tensor. At the linear level,
this can be done by including the term
\begin{equation}
\mathcal{L}\supset \tilde C^{\mu\nu}\tau_{{\rm M}\mu\nu}
\label{lin_matter}
\end{equation}
where
\begin{equation}
\tau_{\mu\nu}=T_{\mu\nu}-\half\eta_{\mu\nu}T_{\rm M}{}^\alpha_\alpha
\end{equation}
is the trace-reversed energy-momentum tensor. As a result, the equations
of motion reduce to the linearized Einstein equation
\begin{equation}
K_{\mu\nu\alpha\beta}\tilde C^{\alpha\beta}\equiv R^L_{\mu\nu}=\tau_{{\rm M}\mu\nu}
\label{cardinal_flat}
\end{equation}
(where we have taken, for now, the values of the Lagrange multipliers
$\lambda_n$ equal to zero).

The total energy-momentum tensor consists not only of
contributions of matter. There is a contribution of the gravitons
themselves as well, quadratic in $\tilde C^{\mu\nu}$.  As a consequence, the
inclusion of a cubic term in (\ref{cardinal_flat}) is required.  This,
in turn, implies a cubic contribution to the energy-momentum tensor,
corresponding to a quartic term in the Lagrangian. This process
continues indefinitely, yielding in the limit the full Einstein-Hilbert
action \cite{kraichnanetal}. 
It has been shown by Deser~\cite{deser} that this ``bootstrap'' process
can be carried out, for general relativity, in one step if one
rewrites the free graviton action in first order (Palatini) form. 

In order to implement this bootstrap procedure for the cardinal theory,
we pass, as a first step, to the trace-reversed tensor $\fC$ as
\begin{equation}
\fC^{\mu\nu}=-C^{\mu\nu}+\half\eta^{\mu\nu}C^\alpha{}_\alpha.
\end{equation}
The kinetic term in (\ref{cardinal_flat}) can be rewritten in
Palatini form as
\begin{equation}
\fC^{\mu\nu}(\Gamma^\alpha_{\mu\nu,\alpha}-
\Gamma_{\mu,\nu})+\eta^{\mu\nu}(\Gamma^\alpha_{\mu\nu}\Gamma_\alpha-
\Gamma^\alpha_{\beta\mu}\Gamma^\beta_{\alpha\nu})
\label{Palatini}
\end{equation}
where the connection coefficients $\Gamma^\alpha_{\mu\nu}$ are to be
considered as independent additional variables.
It can be shown that the bootstrap of the kinetic term then
terminates after one step, yielding
\begin{equation}
\fC^{\mu\nu}(\Gamma^\alpha_{\mu\nu,\alpha}-\Gamma_{\mu,\nu})
+(\eta^{\mu\nu}+\fC^{\mu\nu})(\Gamma^\alpha_{\mu\nu}\Gamma_\alpha-
\Gamma^\alpha_{\beta\mu}\Gamma^\beta_{\alpha\nu})
\end{equation}
which is equivalent to the Einstein-Hilbert lagrangian
\begin{equation}
(\eta+\fC)^{\mu\nu}R_{\mu\nu}(\Gamma).
\end{equation}
From the latter form we conclude that $(\eta+\fC)^{\mu\nu}$ is naturally
interpreted as a curved-space metric density.

The bootstrap procedure can also be applied to the matter interaction
to determine the form of the matter coupling for cardinal gravity.
For this purpose, the interaction (\ref{lin_matter}) is conveniently
expressed in terms of the trace-reversed energy-momentum tensor
$\tau_{{\rm M}\mu\nu}$ for the matter.  This tensor arises by
variation of the Lagrange density $\mathcal{L}_{\rm M}$ for the matter
fields via
\begin{equation}
-\half \tau_{{\rm M}\mu\nu} =
\frac{\delta\mathcal{L}_{\rm M}(\eta\to\psi)}{\delta\psi^{\mu\nu}}
\bigg|_{\psi\to\eta}
\end{equation}
in the usual way.  We can therefore write
\begin{equation}
\mathcal{L}^L_{{\rm M},\fC} = 
-\half \fC^{\mu\nu}
\tau_{{\rm M}\mu\nu} 
\label{mcint}
\end{equation}
for the matter interaction with the cardinal field $\fC^{\mu\nu}$.
Applying the bootstrap procedure can be shown to generate the Lagrange
density
\begin{equation}
\mathcal{L}_{{\rm M},\fC} = \mathcal{L}_{\rm M}\big|_{\eta\to\eta+\fC}.
\label{cbootmlag}
\end{equation}
This expression corresponds to the usual curved-space matter Lagrangian
if we identify (as above) $(\eta+\fC)^{\mu\nu}$ as the metric density.
For example, the bootstrap procedure applied to the flat-space
electromagnetic energy-momentum tensor yields
\begin{equation}
\mathcal L_{EM}=-{1\over4\sqrt{|\eta+\fC|}}(\eta+\fC)^{\alpha\gamma}
(\eta+\fC)^{\beta\delta}F_{\alpha\beta}F_{\gamma\delta}.
\end{equation}

It is interesting to note at this point that the way we implemented the
bootstrap procedure above is not completely unique. Instead of writing the
full trace-reversed tensor $\fC^{\mu\nu}$ in the Palatini form
(\ref{Palatini}), one could  substitute its fluctuations
$\tilde\fC^{\mu\nu}$ (defined analogously to $\tilde C^{\mu\nu}$
above) instead. While this makes no difference in the linearized
expression, when applying the bootstrap to this form one ends up
with a different result, yielding $(\eta+\tilde\fC)^{\mu\nu}$
as the curved-space metric density. Applying this alternative
procedure to the matter coupling yields a curved-space matter
Lagrangian with an explicit reference to the expectation value
$c^{\mu\nu}$, constituting Lorentz-violating terms that can be related
to the Standard Model Extension~\cite{kostelecky-colladay,
kostelecky-gravity}. Its parameters for Lorentz violation have been
subject to numerous experimental measurements~\cite{kostelecky-russell}.
For details see Ref.~\cite{kopo2}.

The most interesting application of the bootstrap is to
the scalar potential V, which we now express as a function of four
scalars $\fX_i$ ($i=1\ldots4$) defined analogously to the $X_i$
defined in Eqs.~(\ref{X1})--(\ref{X4}) but with $C^{\mu\nu}$ replaced by
the trace-reversed tensor $\fC^{\mu\nu}$. As it turns out,
for the procedure to be able to be applied to $V$, the latter
needs to satisfy nontrivial integrability conditions that strongly
restrict its functional form. For instance, it follows that the
unique lowest-order (in terms of the total power of $\fC$) integrable
polynomials are
\begin{eqnarray}
\fV_0&=&1,\nonumber\\
\fV_1&=&\fX_1,\nonumber\\
\fV_2&=&\fX_2-\frac{\fX_1^2}2,\nonumber\\
\fV_3&=&\fX_3-\frac{3\fX_1\fX_2}4+\frac{\fX_1^3}8.
\end{eqnarray}
More generally, it follows that any polynomial obtained as the term
at order $q$ in the series expansion of $\sqrt{|\mbox{det}[1+\fC\eta]|}$
is a solution $\fV_q$. Applying the bootstrap to $\fV_q$ yields
\begin{equation}
\fV_q\to\sqrt{|\mbox{det}[1+\fC\eta]|}-\sum_{k=0}^{q-1}\fV_k.
\end{equation}
We postulate that any integrable potential can be obtained as a suitable
linear combination of the $\fV_q$ polynomials.

Of particular interest are scalar potentials $\fV_\fC$
that have a Taylor expansion of the form
\begin{equation}
\fV_\fC(\{\fX_i\})={1\over2}\sum_{i,j}a_{ij}(\fX_i-\fx_i)(\fX_j-\fx_j)
        +\mathcal{O}(\fX_i-\fx_i)^3
\label{potential-Taylor}
\end{equation}
with $a_{ij}$ real constants. If we impose that $a_{ij}$ be positive
definite, $\fV_\fC$ has a local minimum at
$\fX_i=\fx_i$ ($i=1\ldots4$), and it can represent a possibly stable vacuum.
It can be shown that the integrability conditions impose constraints
on the $a_{ij}$ and on the higher order coefficients, but that
a nontrivial solution space exists~\cite{kopo2}.
We have verified numerically that there is a non-empty subset of positive
definite solutions, confirming the possibility of scalar potential
with stable vacua~\cite{kopo2}.

It is reasonable to expect that, in the limit $a_{ij}\to\infty$ the
integrable potentials $\fV_\fC$ of eq.\ \ref{potential-Taylor}
correspond to integrable potentials
in the linearized limit considered above
\begin{equation}
\fV_\fC^L=\lambda_1(\fX_1-\fx_1)+
\lambda_2\Bigl(\fX_2-\frac{\fX_1^2}2-\fx_2+\frac{\fx_1^2}2\Bigr)+
\lambda_3\Bigl(\fX_3-\frac{3\fX_1\fX_2}4+\frac{\fX_1^3}8
-\fx_3+\frac{3\fx_1\fx_2}4-\frac{\fx_1^3}8\Bigr)
+\ldots,
\end{equation}
fixing the value of $\fX_i$ to be $\fx_i$.
We conclude that in this limit, the bootstrapped cardinal model 
corresponds to General Relativity in the gauge defined by the
constraints $\fX_i=\fx_i$. This is also the case for a potential
with finite $a_{ij}$, if we only consider energy scales low enough
such that any fluctuations away from the minimum in the potential
can be neglected.

\section{Vacuum energy-momentum tensor}

After applying the bootstrap procedure described above we end up
with a Lagrangian density of the form
\begin{equation}
(\eta+\fC^{\mu\nu})R_{\mu\nu}(\Gamma)-\sqrt{|\eta+\fC|}\,V(\fX_1,\fX_2,\fX_3,\fX_4)
 + \mathcal L_M(C,\eta,\phi_i,\partial_\mu\phi_i)
\end{equation}
The linearized equations of motion become:
\begin{equation}
K_{\mu\nu\alpha\beta}\fC^{\alpha\beta}=(\eta_{\mu\nu}\partial_1+
2\eta_{\mu\alpha}c^{\alpha\beta}\eta_{\beta\nu}\partial_2+...)V+
\tau_{\mu\nu}^{(m)}(\eta,\phi_i,\partial_\mu\phi_i)
\label{full_lin_eqs}
\end{equation}
where we defined
\begin{equation}
\partial_n\equiv{\partial\over\partial\fX_n}\qquad (n=1\ldots4).
\end{equation}

We see from Eq.\ (\ref{full_lin_eqs}) that the first term on the
right-hand side naturally takes the form of a (trace-reversed)
energy momentum tensor. Explicitly, we can identify a ``vacuum
energy-momentum tensor'' 
\begin{equation}
T_{\mu\nu}^{(vac)}=\tau_{\mu\nu}^{(vac)}
-{1\over2}\eta_{\mu\nu}\bigl(\tau^{(vac)}\bigr)^\alpha_\alpha
\label{vacemtensor}
\end{equation}
with
\begin{equation}
\tau_{\mu\nu}^{(vac)}=(\eta_{\mu\nu}\partial_1+
2\eta_{\mu\alpha}c^{\alpha\beta}\eta_{\beta\nu}\partial_2+...)V.
\end{equation}
It takes nonzero values whenever the scalar potential takes values
away from the minimum.

Explicit solutions of the linearized
equations of motion can be obtained with nonzero vacuum
energy-momentum tensor. In the latter case,
independent initial/boundary field values can be defined on maximally
4 suitably defined timelike/spacelike spacetime slices.  If the matter
energy-momentum tensor is known to be independently conserved (e.g.,
by symmetry arguments), the same has to be true for the vacuum
energy-momentum tensor.  In such a case, choosing $T_{\mu\nu}^{(vac)}$
to be zero at the initial value spacetime slices ensures it is zero
throughout spacetime.

\section{Conclusions}
We showed it is possible to construct an alternative theory of gravity, the
cardinal model, based on spontaneous
breaking of Lorentz symmetry.
The massless gravitons can be interpreted as Nambu-Goldstone modes
of the spontaneously broken Lorentz symmetry.  The full nonlinear
form of the Lagrangian is fixed by consistent coupling of gravity to
the total energy-momentum tensor, and can be constructed by a
bootstrap process, starting with an initial Lagrangian for the cardinal
field consisting of
a quadratic kinetic term and a scalar potential.
As it turns out, consistency of the bootstrap process imposes strong
restrictions on the form of the scalar potential.
Nevertheless, consistent potentials exist with local minima.
At low energy, the Lagrangian
corresponds to the Einstein-Hilbert action, with the possible presence
of a nonzero ``vacuum energy-momentum tensor''.
At high energy, four extra massive graviton modes appear that modify the
dynamics of the theory.
An open problem remains the classification of all possible bootstrapped
potentials and a study of their properties.
Other issues that merit further study are the effect of the massive modes,
in particular near singularities or at high temperatures, as well as a study
of the quantization of the cardinal model.

\section*{Acknowledgments}
It is a pleasure to thank Alan Kosteleck\'y for collaboration.
Financial support by the Funda\c c\~ao para a Ci\^encia e a Tecnologia
is gratefully acknowledged.

\end{document}